\documentclass[sigconf]{acmart}

\usepackage{graphicx}
\usepackage{amsmath}
\usepackage{booktabs}
\usepackage{algorithm}
\usepackage{algorithmic}
\usepackage{amsfonts}
\usepackage{multirow}
\usepackage{makecell}
\usepackage{subfigure}
\usepackage{color}
\usepackage{bm}
\usepackage{epstopdf}
\usepackage{url}
\usepackage[cal=cm]{mathalfa}
\usepackage{balance}
\usepackage{threeparttable}
\usepackage{lipsum}
\usepackage{enumitem}
\usepackage{pifont}
\usepackage{tabularx}
\usepackage{makecell}
\usepackage{float}
\usepackage[table]{xcolor}
\usepackage{array}
\usepackage{afterpage}

\newcommand{\myorange}[1]{\textcolor{orange}{#1}}  

\setlist[itemize]{leftmargin=*}

\AtBeginDocument{%
  \providecommand\BibTeX{{%
    \normalfont B\kern-0.5em{\scshape i\kern-0.25em b}\kern-0.8em\TeX}}}

\begin{document}
\title{PushGen: Push Notifications Generation with LLM}

\author{Shifu Bie}
\affiliation{
  \institution{Kuaishou Technology}
  \city{Beijing}
  \country{China}
}
\email{bieshifu03@kuaishou.com}

\author{Jiangxia Cao}
\authornote{Jiangxia Cao and Guorui Zhou are Corresponding authors.}
\affiliation{
  \institution{Kuaishou Technology}
  \city{Beijing}
  \country{China}
}
\email{caojiangxia@kuaishou.com}

\author{Zixiao Luo}
\affiliation{
  \institution{Kuaishou Technology}
  \city{Beijing}
  \country{China}
}
\email{luozixiao@kuaishou.com}

\author{Yichuan Zou \\  Lei Liang \\ Lu Zhang}
\affiliation{
  \institution{Kuaishou Technology}
  \city{Beijing}
  \country{China}
}
\email{zouyichuan, lianglei03}
\email{zhanglu14@kuaishou.com}

\author{Linxun Chen \\ Zhaojie Liu \\ Xuanping Li}
\affiliation{
  \institution{Kuaishou Technology}
  \city{Beijing}
  \country{China}
}
\email{chenxi36, zhaotianxing}
\email{lixuanping@kuaishou.com}

\author{Guorui Zhou* \\ Kaiqiao Zhan \\ Kun Gai}
\affiliation{
  \institution{Kuaishou Technology}
  \city{Beijing}
  \country{China}
}
\email{zhouguorui,zhankaiqiao@kuaishou.com}
\email{kun.gai@qq.com}

\renewcommand{\shortauthors}{Shifu Bie et al.}

\copyrightyear{2026}
\acmYear{2026}
\setcopyright{cc}
\setcctype{by}
\acmConference[WSDM '26]{Proceedings of the Nineteenth ACM International Conference on Web Search and Data Mining}{February 22--26, 2026}{Boise, ID, USA}
\acmBooktitle{Proceedings of the Nineteenth ACM International Conference on Web Search and Data Mining (WSDM '26), February 22--26, 2026, Boise, ID, USA}
\acmPrice{}
\acmDOI{10.1145/3773966.3779373}
\acmISBN{979-8-4007-2292-9/2026/02}

\renewcommand{\shorttitle}{PushGen}

\begin{abstract}
We present \textbf{PushGen}, an automated framework for generating high-quality push notifications comparable to human-crafted content.
With the rise of generative models, there is growing interest in leveraging LLMs for push content generation.
Although LLMs make content generation straightforward and cost-effective, maintaining stylistic control and reliable quality assessment remains challenging, as both directly impact user engagement. 
To address these issues, PushGen combines two key components: 
(1) a controllable category prompt technique to guide LLM outputs toward desired styles, and (2) a reward model that ranks and selects generated candidates. Extensive offline and online experiments demonstrate its effectiveness, which has been deployed in large-scale industrial applications, serving hundreds of millions of users daily.
\end{abstract}



\maketitle

\section{Introduction}

\myorange{\textbf{Background.}} Push notifications play a crucial role in maintaining daily active users (DAU) and re-engaging users~\cite{zhao2018notification, hindman2015personalization}. On platforms such as \textbf{Kuaishou}, \textbf{TikTok} and \textbf{Douyin}, daily volumes range from millions to billions~\cite{yancey2020sleeping, yue2022learning, kroer2023fair}. 
Notification quality directly affects CTR and ecosystem health, with low-quality notifications often leading users to disable pushes or even uninstall the app~\cite{wang2025beyond}.

In industry, two main notification pipelines are deployed~\cite{yang2025llm, zeng2024large}:

\begin{itemize}
    \item \textbf{Template-based notifications (private domain)}: These target videos from creators with whom users have strong behavioral connections(e.g., those they follow or frequently interact with), and the content quality tends to be stable.
    \item \textbf{Dense-content notifications (public domain)}: These target a small, high-quality pool of \emph{public} videos, requiring deeper content understanding to generate engaging notifications, which is more challenging and costly.
\end{itemize}
\emph{Manual} generation of public-stream push notifications is limited in scalability, as only a few thousand items can be written per day, while daily video uploads reach into the hundreds of millions to billions. Moreover, only a small fraction (\(\sim 0.2\%\)) of public videos are suitable for pushing, making automating high-quality content generation a critical challenge.

\begin{figure}[t!]
  \centering
  \includegraphics[width=9cm,height=5cm]{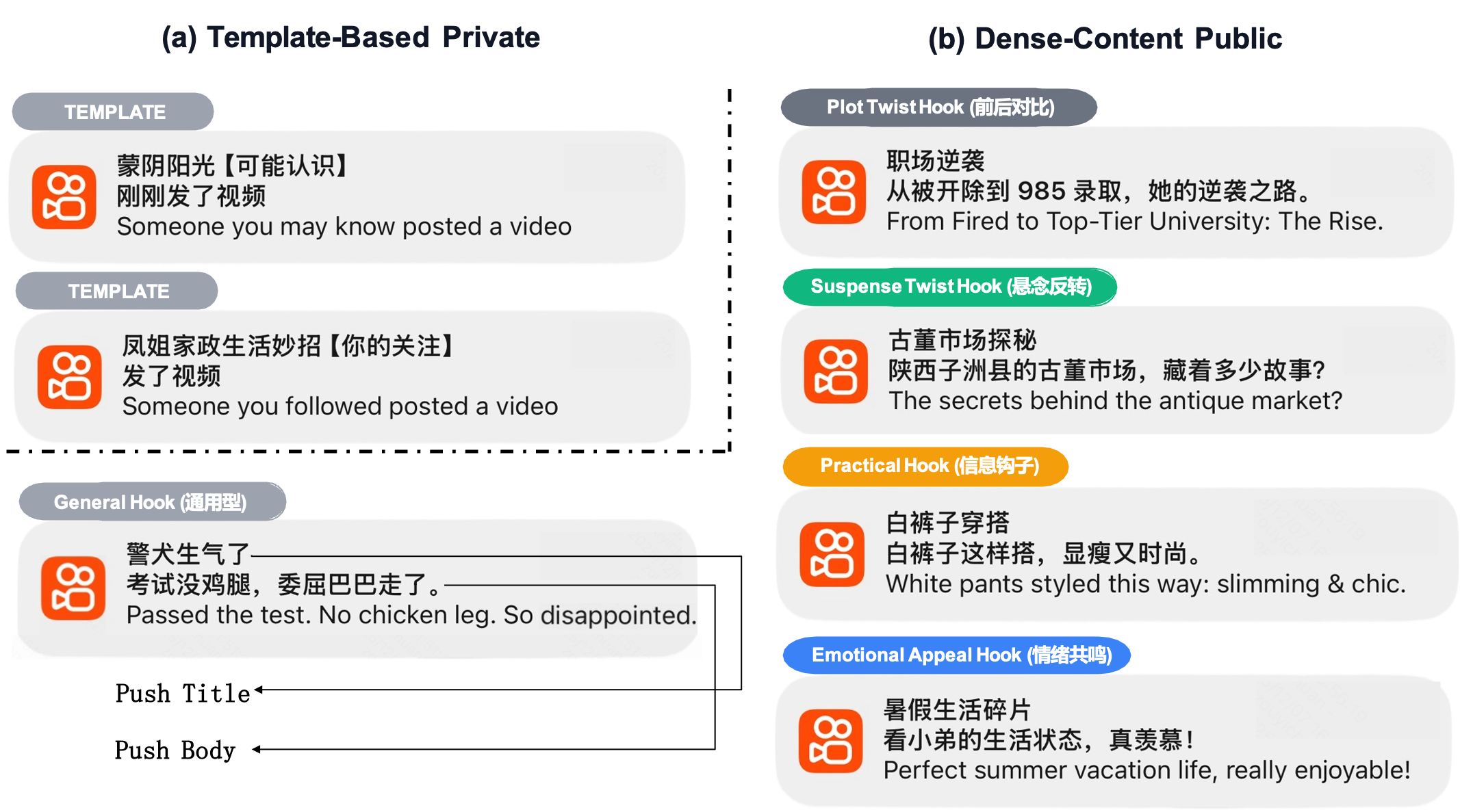}
  \caption{Push Notifications at our Application.}
  \label{push_notifications}
\vspace{-0.5cm}
\end{figure}

\begin{figure*}[t!]
  \centering
  \includegraphics[width=15cm,height=6.5cm]{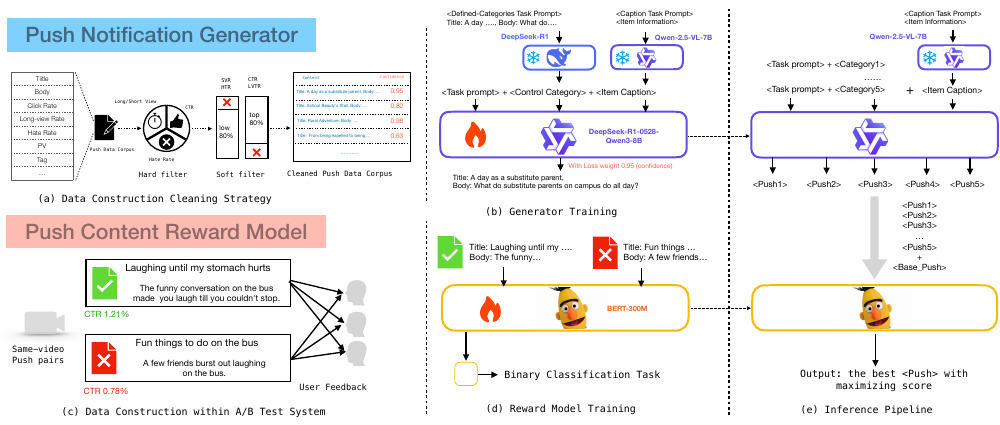}
  \caption{The overall framework of PushGen, where the \texttt{<Other\_Push>} indicates the DeepSeek V3 671B + RAG Push results.}
  \label{push}
  \vspace{-0.3cm}
\end{figure*}

\myorange{\textbf{Related Works.}} At Kuaishou, two methods have been deployed:
\begin{itemize}
\item \textbf{Small LLM with Supervised Fine-Tuning (SFT)}: Fine-tuning a pretrained model (e.g., \emph{Dense Qwen-14B}~\cite{bai2023qwen}) on in-house data, producing human-like content and supporting large-scale public-stream mining since early 2024.

\item \textbf{Large LLM with Retrieval-Augmented Generation (RAG)}: Using stronger \emph{MoE} models (e.g., \emph{DeepSeek V3 671B}~\cite{liu2024deepseek}) with retrieval-augmented signals~\cite{chen2025ctr}, deployed in mid-2025 with improvements in DAU \(+0.052\%\) and CTR \(+4.330\%\).
\end{itemize}

Despite these advances, two \emph{core gaps} remain:
(i) \textbf{Style Control:} SFT lacks reliable control over style/theme during inference.
(ii) \textbf{CTR Alignment:} There is no reliable ex-ante metric to predict which content is more likely to attract clicks, complicating pre-launch candidate selection.

\myorange{\textbf{Motivation.}} We aim to tackle two key challenges:
\begin{itemize}
\item \textbf{Controllable diversity}: As shown in Figure~\ref{push_notifications}, the system must generate diverse candidates within predefined styles(e.g., \emph{suspense, emotional, practical, etc}), maintaining freshness and avoiding homogeneity. Current SFT lacks the necessary style conditioning for this.
\item \textbf{An ex-ante evaluator aligned with CTR}: Point-wise scoring is often unreliable due to factors such as creator popularity, topical relevance, intrinsic video quality, and audience composition, causing misalignment with actual CTR. A more effective approach is needed. We propose using small-traffic A/B tests to generate paired comparisons and train a pairwise reward model (RM) that ranks candidates pre-launch.\end{itemize}

\myorange{\textbf{Our Contribution.}} We propose \textbf{PushGen}, a framework for automated push generation that addresses these challenges.
\begin{itemize}
  \item \textbf{Category-controllable diverse generation:} We introduce a data-cleaning and weighting pipeline to distill high-quality samples, enabling the generator to reliably produce diverse candidates across various styles.
  \item \textbf{Pairwise reward modeling:} We train a pairwise RM~\cite{schulman2017proximal} using A/B traffic to rank candidates based on CTR differences, directly targeting the deployment objective, unlike point-wise estimation.
  \item \textbf{At-scale deployment and gains.} PushGen is deployed at scale, serving hundreds of millions of users. Our online A/B tests show a \textbf{+14.080\%} CTR increase, with  \textbf{85.2\%} replacement by matching-generated content, a \textbf{+4.026\%} CTR improvement in the public domain, and a \textbf{+0.067\%} increase in unique DAU.
  \end{itemize}

\section{PushGen Workflow}
As illustrated in Figure~\ref{push}, the PushGen workflow consists of two main stages: \textbf{Push Notification Generator:} Dense captions are generated from video frames and ASR data, then combined with the video original title, topics, and platform category. Diverse candidate notifications are generated through fine-tuning LLMs. \textbf{Push Content Reward Model:} A pairwise reward model ranks candidate notifications based on CTR differences, ensuring the selection of optimal notifications aligned with user preferences.
PushGen enables "controllable diversity in candidate generation" through the Push Notification Generator, and "CTR-aligned selection" through the Push Content Reward Model. 
High-quality notifications are replaced our former DeepSeek V3 671B + RAG Push results.

\subsection{Push Notification Generator}

\subsubsection{VLM Dense Caption}
Given training/inference resource and QPS constraints, we adopt a two-stage pipeline: a frozen vision-language model (VLM, 7B) generates an \emph{explicit} textual description of the video, and a text LLM generates the push copy. This reduces computational cost of end-to-end multimodal decoding while ensuring an auditable, traceable semantic basis for factual consistency. Inputs include video frames, ASR transcripts, the original title, and category metadata, with the output being a dense caption that serves as content evidence for the downstream generator.

\subsubsection{Dataset Design (Filtering and Weighting)}
To mitigate sample polarization and noise, we use a three-step strategy:
\paragraph{(1) Statistical hard filter}
    \begin{equation*}
    \footnotesize
    \begin{split}
    \texttt{filter}_{\texttt{hard}} = (\texttt{CTR}>0.6\%) \& (\texttt{SVR}<40\%) \&(\texttt{LVTR}>50\%) \& (\texttt{HTR}<1\%) \& (\texttt{PV}>800)
    \end{split}
    \end{equation*}
\noindent where CTR/SVR/LVTR/HTR represent click/short-view/long-view/hate rates of user feedback, and \textbf{PV} denotes exposure volume.
\paragraph{(2) Tag-wise soft filter}
To reduce category/cluster bias, we perform within-cluster quantile cropping:
\begin{equation*}
\small
\begin{split}
\texttt{filter}_{\texttt{soft}} = 
\begin{cases} 
\texttt{CTR}\&\texttt{LVTR}\ \ \text{out of the \textbf{bottom} 20\% in each tag cluster}  \\
\texttt{SVR}\&\texttt{HTR}\ \ \text{out of the \textbf{top} 20\% in each tag cluster}  \\
\end{cases} \\
\end{split}
\end{equation*}
\paragraph{(3) Confidence weighting}
On the cleaned corpus, we assign sample-specific weights based on quality and exposure:
\[
\small
\text{confidence} = 0.3 + 0.35 \cdot \frac{\min(\text{CTR}, 0.1)}{0.1} + 0.35 \cdot \log\left(\frac{\min(\text{PV}, 10000)}{10000}\right)
\]

\subsubsection{SFT Training}
We explicitly inject \emph{category controllability} into training/inference for stable style coverage/generation.

\paragraph{(1) Control Category (style condition)}
A large frozen LLM (e.g., DeepSeek-R1~\cite{guo2025deepseek}) classifies push samples into predefined styles (e.g., \emph{Suspense, Emotion, Practical \dots}) after multiple consistent inquiries, producing \(\langle\text{Control Category}\rangle\).
\begin{equation*}
\footnotesize
\begin{split} 
\texttt{<Control Category>}& = \texttt{LLM}_{\texttt{Category}}^{\texttt{Frozen}}(\texttt{<Defined-Categories Task Prompt>}\texttt{<Label\_Push>}) \\
\end{split}
\end{equation*}

\paragraph{(2) Push content generation (SFT)}
Using ``task instruction + style condition + content evidence'' as input, we fine-tune the text LLM to generate content aligned with the selected category:
\begin{equation*}
\small
\begin{split}
\texttt{<Push>} = &\texttt{LLM}_{\texttt{Gen}}(\texttt{<Task Prompt>}\texttt{<Control Category>}\texttt{<Item Caption>})) \\
&\mathcal{L}_{\texttt{sft}} = \texttt{confidence} \times \texttt{SFT}(\texttt{<Label\_Push>}, \texttt{<Push>})
\end{split}
\end{equation*}
where \(\langle \text{Item Caption}\rangle\) is the \textbf{VLM dense caption} described in Section~2.2.1.  

During inference, we generate multi-style candidate copies by conditioning on predefined categories and applying mild sampling techniques(e.g., top-$p$, temperature, repetition penalty), then select the best candidates by the reward model.

\subsection{Push Content Reward Model}
Based on the fine-tuned model, we can generate multiple contents variations with diverse styles. This section focuses on the reward model that selects the most reliable content.

\subsubsection{Data Pipeline}
We construct a supervised dataset by using our A/B testing platform, distributing different push notifications for the same short video to a small portion of online traffic.
This enables us to measure user preferences and acceptance for varying content, as shown in Figure~\ref{push}(c).

\subsubsection{Model Tuning/Inference}
We experimented with both point-wise and pair-wise learning paradigms for CTR estimation.
However, point-wise learning was unsuitable due to confounding factors such as author popularity, topical alignment, and video quality. Instead, we focus on relative CTR comparisons.
For a push pair $\langle \texttt{Push}_1, \texttt{Push}_2 \rangle$, the input is concatenated as:
\[
\small
[\texttt{CLS}] \; \texttt{Push}_1 \; [\texttt{SEP}] \; \texttt{Push}_2 \; [\texttt{SEP}]
\]
The reward model is defined as:
\[
\small
\mathbf{h}_{\texttt{[CLS]}} = \texttt{BERT}(\langle \texttt{Push}_1, \texttt{Push}_2 \rangle), \quad
\mathbf{r} = \sigma(\texttt{MLP}(\mathbf{h}_{\texttt{[CLS]}})),
\]
with the training objective
\[
\small
\mathcal{L}_{\texttt{reward}} = \texttt{BCE}(\mathbf{r}, \texttt{label}),
\]
where $\mathbf{h}_{\texttt{[CLS]}} \in \mathbb{R}^d$ is the [CLS] embedding, $\mathbf{r} \in (0,1)$ the predicted probability that $\texttt{Push}_1$ outperforms $\texttt{Push}_2$, and $\texttt{label}\in\{0,1\}$ indicates the ground truth (1 if $\texttt{CTR}(\texttt{Push}_1) > \texttt{CTR}(\texttt{Push}_2)$, 0 otherwise). We use BERT as backbone for the reward model, which is well-suited for discrimination tasks.

\section{Discussion of Reward Hacking}
Reinforcement learning (RL) has recently played a pivotal role in advancing LLMs, enhancing reasoning abilities in models like DeepSeek-R1 and Qwen3. In PushGen, we explored both preference optimization (e.g., DPO~\cite{rafailov2023direct}) and RL-based techniques (e.g., PPO~\cite{schulman2017proximal}, GRPO~\cite{guo2025deepseek}) to link the SFT model with the reward model.

In push notification generation, user click feedback (CTR) is highly sensitive to entity salience: such as mentions of idol names or specific numbers, which often substantially increase click probability. This reflects natural preferences and attention allocation. However, when RL/PO dives the generator, such salience biases in the reward signal can be amplified, leading the model to learn reward shortcuts instead of aligning with video facts and text quality. This can result in content hallucination and factual inconsistency, which is unacceptable in the push content scenario.

Based on these observations, we currently rely on SFT + pairwise RM for controllable generation and relative selection, while reserving RL/PO for future iterations. We plan to further incorporate content legality checks to constrain the reward signal.

\begin{figure}[t!]
\begin{center}
\includegraphics[width=8cm]{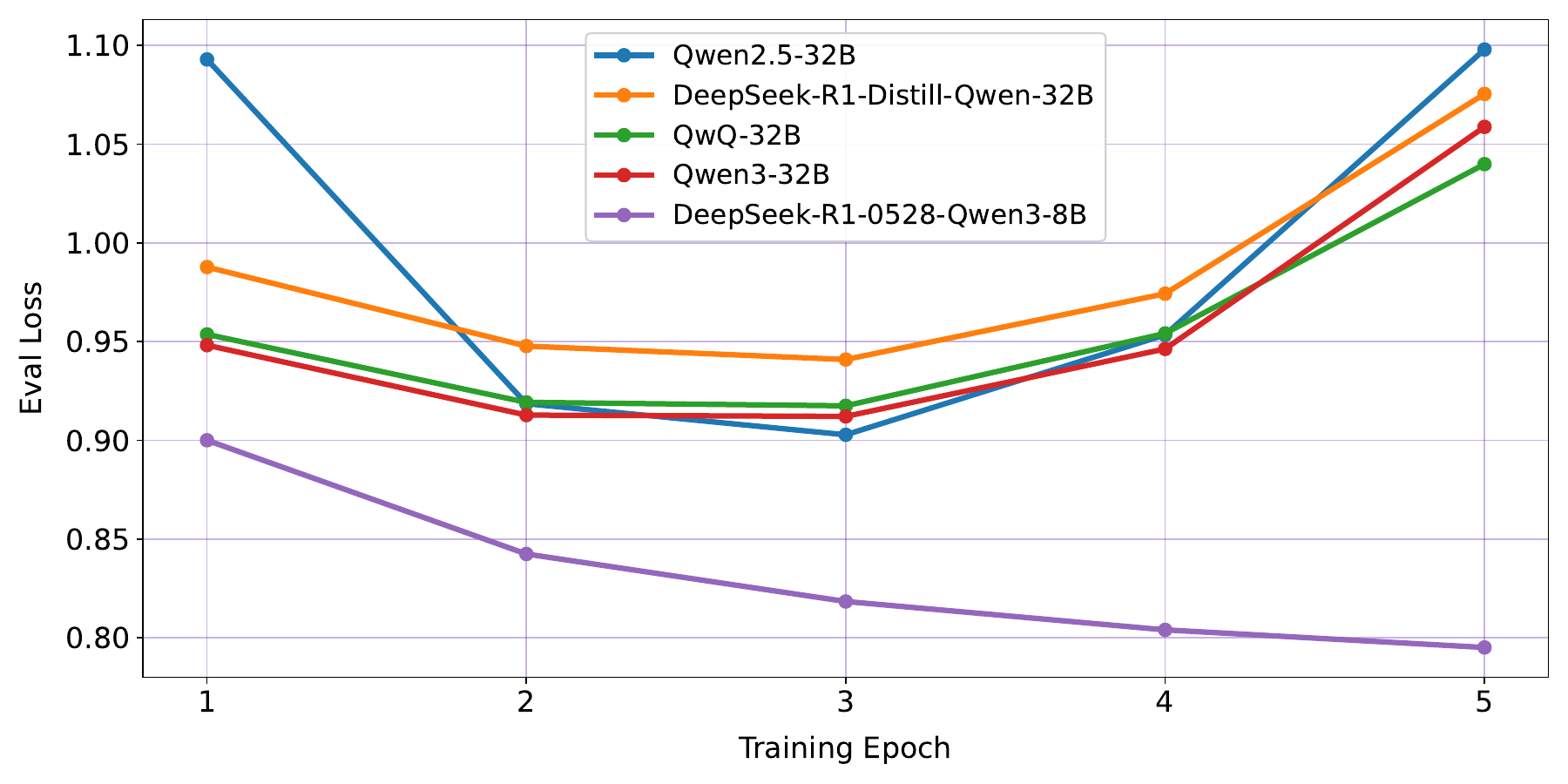}
\caption{Loss curves of different LLMs during training.}
\label{sftmodel}
\end{center}
\end{figure}

\begin{table}[t!]
\centering
\caption{Offline results(\%) in terms of Accuracy.}
\setlength{\tabcolsep}{7pt}{
\begin{tabular}{lcccc}
\toprule
\multirow{2}{*}{Sample Group} & \multicolumn{2}{c}{BERT}  & \multicolumn{2}{c}{Qwen} \\
\cmidrule(r){2-3} \cmidrule(r){4-5} & 324M & 710M & 0.6B & 1.5B \\
\hline
0\%-25\% & 70.03\% & 68.65\% & 66.45\% & 67.92\% \\
25\%-50\%  & 71.02\% & 72.01\% & 68.19\% & 69.99\% \\
50\%-75\%  & 74.05\% & 76.14\% & 72.32\% & 75.50\% \\
75\%-100\%  & 79.24\% & 78.30\% & 76.49\% & 78.92\% \\
Overall  & 73.58\% & 73.77\% & 70.77\% & 73.00\% \\
\bottomrule
\end{tabular}
}
\label{rewardmodel}
\end{table}

\begin{figure}[t!]
\begin{center}
\includegraphics[width=8cm]{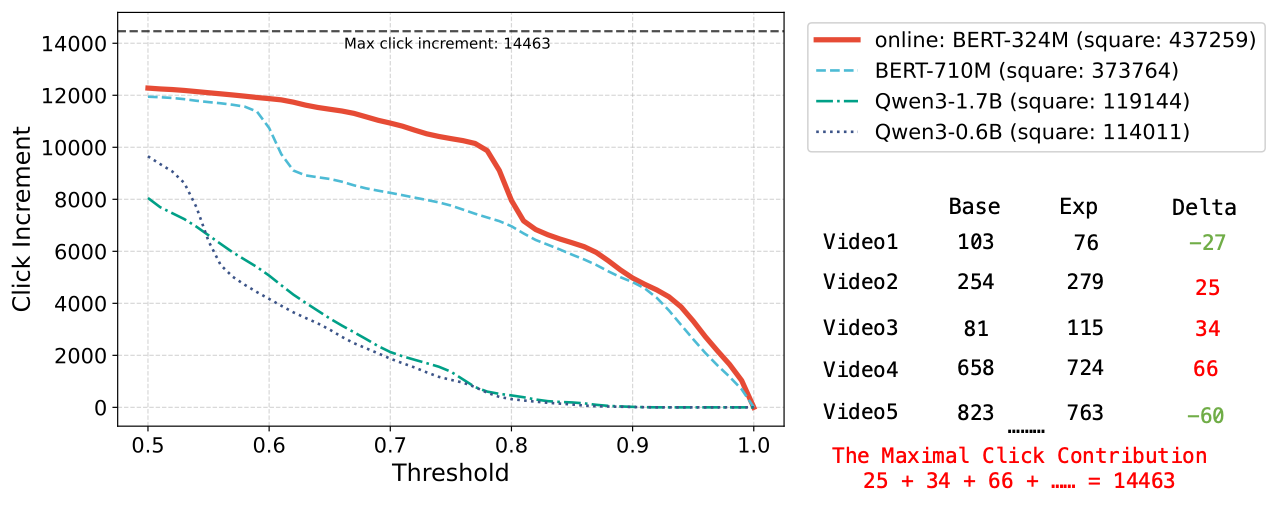}
\caption{Reward model classification ability analysis.}
\label{reward_analysis}
\end{center}
\end{figure}

\begin{table}[t!]
\small
\centering
\caption{Content Style Distribution Before and After SFT.}
\setlength{\tabcolsep}{1pt}{
\begin{tabular}{lccccccc}
\toprule
\multirow{2}{*}{Generator} & \multirow{2}{*}{Base} & \multicolumn{6}{c}{Replacement} \\
\cmidrule(r){3-8} & & Suspense &Emotion &Praticical &Plot &General & Other \\
\hline
Pre-train &56.0\%& 6\%& 5.6\%&1.8\% & 8.8\% &5.8\% &16\% \\
 SFT &17.0\% &11.8\%& 10.8\%& 9\%& 13.6\%& 11.4\% & 26.4\% \\
\bottomrule
\end{tabular}
}
\label{diversity}
\end{table}

\begin{table}[ht!]
\small
  \centering
  \begin{threeparttable}
  \caption{Online A/B testing results in Push service.}
  \label{tab:ab}
  \begin{tabular}{lccc}
    \toprule
    Metrics & \makecell{Overall\\User}  & \makecell{Mid Active\\User} & \makecell{Low Active\\User}  \\
    \midrule
    DAU (Dedup.\footnotemark) & +0.067\% & +0.196\% & +0.176\% \\
    CTR (Public) & +4.026\% &  +3.034\% & +2.600\% \\
    CTR (Replaced) & +14.080\% & - & - \\
    PV (Replaced)  & 85.2\% & - & - \\
    \bottomrule
\end{tabular}
  \end{threeparttable}
\end{table}

\section{Experiments}
In this section, we conduct extensive online and offline experiments to verify the efficacy of our PushGen approach.

\afterpage{
\footnotetext{DAU (Dedup.): Daily Active Users deduplicated across clients; CTR (Public): Click-Through Rate on the combined pool of machine- and human-generated push content; CTR (Replaced): Click-Through Rate on machine-generated content replaced by our system; PV (Replaced): the send volume by replaced push content.}
}

\subsection{Offline Performance}
During offline iteration, we first constructed a cleaned training corpus of over 70k samples selected from a historical million-scale push notification repository. Subsequently, for SFT, we experimented with various models for data fitting, including text-only reasoning models, multimodal reasoning models, 8B dense models, and 32B mixture-of-experts (MoE) architectures.
As shown in Figure~\ref{sftmodel}, the larger 32B models exhibited pronounced overfitting tendencies across multiple epochs, while the 8B model maintained robust generalization across extended training iterations.
Therefore, we adopted DeepSeek-R1-0528-Qwen3-8B as the base model for SFT.

For the reward model, we evaluated several small models (< 2B parameters), reporting both overall prediction accuracy and stratified accuracy across four difficulty levels defined by CTR differences between positive and negative sample pairs. As shown in Table~1, encoder-based BERT models consistently outperformed decoder-based GPT models, likely due to the latter’s higher data requirements for comparable discriminative ability under the reward modeling paradigm. We therefore adopted BERT as the base architecture for reward model training.

\subsection{Ablation Study}

\textbf{Reward Model Analysis.} To further evaluate the capability of the reward model, we conducted an ablation study.
Two push notification variants from the same short video pool were allocated to approximately balanced traffic exposure, allowing us to collect real user click feedback.
The relative CTR differences naturally revealed which content performed better or worse.
For each short video, we compared the diverse candidate push notifications generated by the SFT model with the corresponding baseline push notification.
The reward model was then used to identify the better option, and the selected content was served as the Exp push for that video. 
Ideally, if the reward model always selected the ground-truth higher-CTR content for the Exp group, we could reach a theoretical upper bound on performance, indicating maximal click increment.(e.g., +14,463 clicks, as illustrated in Figure~\ref{reward_analysis}).
Specifically, $x$ denotes the probability that the Exp push outperforms the Base push, defined as
\[
x = \frac{1}{1 + e^{-(s_{\text{Exp}} - s_{\text{Base}})}},
\]
where $s_{\text{Exp}}$ and $s_{\text{Base}}$ are the reward model scores of the Exp and Base content, respectively. Meanwhile, $y$ represents the total click increment of Exp group relative to the Base, aggregated over all videos with predicted probability exceeding $x$.
In practice, performance approached but did not reach this bound.
As shown in Figure~\ref{reward_analysis}, although BERT-700M slightly outperformed BERT-300M in offline evaluation (73.77\% vs. 73.58\%), the 300M model demonstrated superior online performance with a flatter curve and larger area under the curve. 
Consequently, we adopted BERT-300M as the base reward model for training in PushGen framework.

\textbf{SFT Model Analysis.} Building on the trained reward model, we further evaluated the generation quality and diversity of PushGen.
We compared outputs from a raw pre-trained LLM and our SFT-enhanced LLM using the same inputs and prompts. For each video, both models generated content in different styles, and the reward model selected the highest-scoring one. The distribution of these selected styles among all videos is summarized in Table~\ref{diversity}.
Results show that:
(1) After SFT, our 8B model substantially increased the proportion of generated contents outperforming the baseline, rising from 44\% to 83\%, effectively replacing the push content process originally handled by the base model (671B DeepSeek-V3 with RAG);
(2) controllable category generation significantly enhances the push content diversity, e.g., from 6\% to 11.8\% in suspense twist hook, 1.8\% to 9\% in practical hook and so on.

\subsection{Online Performance}
We deployed it in production and conducted multi-day online A/B testing.
As reported in Table~\ref{tab:ab}, PushGen consistently delivered online gains.
CTR on modified machine-generated content increased by +14.08\%, with baseline copies replaced across 85.2\% of push send volume.
In the public domain, CTR improved by +4.026\%. 
Moreover, deduplicated DAU rose by +0.067\% overall, with gains of +0.196\% and +0.176\% for mid- and low-active user groups, respectively.

\section{Conclusion}
In this paper, we presented a complete solution for push notification generation, covering both data construction and model design.
Specifically, we introduced a category-controllable push notification generator that ensures diversity and controllability of the generated content.
We further described a reward model designed to identify and select push content that better aligns with user preference.
Looking forward, we plan to extend this work in the following directions: (i) leverage vision-language models (VLMs) as the generator backbone to produce more accurate content, (ii) develop reinforcement learning strategies tailored to connect the SFT model with the reward model more effectively.
\newpage

\balance
\bibliographystyle{ACM-Reference-Format}
\bibliography{sample-base-extend.bib}
\end{document}